\documentclass{article}
\usepackage{graphicx}
\usepackage{epsfig}

\newcommand{\er}{\end{eqnarray}}
\newcommand{\br}{\begin{eqnarray}}
\newcommand{\be}{\begin{equation}}
\newcommand{\ee}{\end{equation}}
\newcommand{\epe}{\end{equation}}
\newcommand{\bea}{\begin{eqnarray}}
\newcommand{\eea}{\end{eqnarray}}
\newcommand{\ba}{\begin{eqnarray}}
\newcommand{\ea}{\end{eqnarray}}
\newcommand{\epa}{\end{eqnarray}}

\begin{document}

\begin{center}
{\huge{Fadeev Jackiw Analysis of Constraints \\ on Noncomutative Hall Effect}}
\end{center}

\centerline{\bf {C.F.L. Godinho} \footnote{godinho@if.ufrj.br, godinho@cbpf.br}}
\begin{center}
\normalsize Centro Brasileiro de Pesquisas Fisicas (CBPF),LAFEX
\newline
Rua Dr. Xavier Sigaud, 150 - Urca 22290-180 - Rio de Janeiro -
Brasil
\end{center}
\begin{abstract}Constrained systems are fundamental to understanding of several physical realities.  Even so the Hall effect is one of more revisited issue we can still find new approaches to obtain old and new important relations. 
In this paper a semi classical formulation is considered where an Chern-Simons gauge invariant theory is constructed for a Schroedinger field.  The main idea is to describe both classical and integer Hall effect.  
We build up the constraints this model by means of the Fadeev-Jackiw quantization algorithm.  In a second step we consider a noncommutative extension to the action.  In this extended approach noncommutative constraints relations are obtained and guide us to an interesting adjustment factor on the conductivity expression.  
\end{abstract}

\vfill\eject

\section{Introduction}
The quantum Hall \cite{Klitz} effect has been one of the main objects under study for theoretical and experimental physics.
\\
During the last few years we can observe substantial progress, in both areas of research , nowadays it is possible to apply some mathematical methods more precisely the noncommutative geometry concepts to deepen the knowledge about this issue.
The quantum Hall system has two dual Chern-Simons descriptions, associated with 
its hydrodynamic and statistics respectively.  The use of CS terms in the action has been 
strongly observed to describe systems where this phenomenon occurs.  Recently Susskind \cite{Sus}presented some interesting arguments that the hydrodynamic CS theory has a noncommutative gauge symmetry.
During the last 10 years a lot of works have been devoted to
understanding the several implications of noncommutative geometry
in physics where now the space-time is deformed in relation to an
ordinary space-time and its coordinates obey the well known 
relation $[ x_{\mu}\,,\, x_{\nu}]\,=\,i \theta_{\mu \nu}$
where $\theta_{\mu \nu}$ is real and antisymmetric constant.
The first ideas about space-time non-commutativity were formulated by Heisenberg \cite{Hei} in the thirties, although the first published work on the subject appeared in 1947 \cite{Sney}.

Our proposal is to understand what kind of contribution we may obtain researching the constraints of a  semi classical model describing  the Hall effect.  In this sense we will apply the Fadeev-Jackiw \cite{FJ,BaWo} algorithm and study the constraints of the classical and quantum Hall \cite{Hall,Klitz}, effect based on the Chern-Simons action coupled to the traditional Schrodinger action.  Basically we intend to implement a  mathematical physics analysis to the constraints on the usual and on the extended model showing natural connection between the Hall conductivity and it.  Then we deal with a noncommutative extension for the model and we will show that it is possible to express
the new Hall conductivity corrected by means of $\theta$ powers.  Our goal is to
show that a new constraint relation arises naturally from the
treatment and gives us an interesting relation to Hall
conductivity where a noncommutative sector shows itself with
particular importance.  Subsequently we analyze its contribution to physical phenomena.

It is also important to remember the similarity between the Fadeev-Jackiw and Dirac \cite{FJ,Dirac} quantization procedures.
It is obvious that the same results could be obtained considering the Dirac approach.  However 
we must consider that the constraints have not the same classification, but their physical nature is the same.
In fact we can say that the constraints in Fadeev-Jackiw's procedure are constraints to Dirac's procedure.  But the opposite consideration is not true always.  The constraints in the Fadeev-Jackiw approach are sometimes called true constraints.

\section{The Model}
Let us start considering an usual model based in a classical action
where the mechanics of a non-interacting
particle system is described on the first term 
by the usual Schroedinger Lagrangian coupled with an electromagnetic potential ($a_m$) represented here by the Chern-Simons (CS) action.     
\footnote{Here the
convections are $\{m,n\}\,=\,1,2$,
$\epsilon_{mn}\,=\,-\epsilon_{nm}\,=\,1$ and $\{\alpha,\beta,
\gamma\,=\,0,1,2$\}} on $(2+1)$ dimensional manifold
$M\,=\,\Sigma\, \times \,R$ was presented before in the paper
\cite{Ghab} and it can be read by
 \bea
\label{Action}
S^{(0)}&=&{1\over{8\pi}}\int dt \quad \Biggl[\int_{\Sigma}
\psi^{\ast}[i\hbar
\partial_t\,-\,{1\over{2m}}(i\hbar
\partial_n\,-\,ea_n)^2\,+ \nonumber \\
&-&ea_0]\psi\,+\,h.c.\,-\,k \int_M \epsilon^{\alpha
\beta \gamma}a_{\alpha}\partial_{\beta}a_{\gamma}\Biggr] \quad.
\eea
A constant of normalization is considered for this action (k) by
considering that it can be treated like a locally constant and
dimensionless parameter.

It is easy to see that the Euler Lagrange equation for $a_m$ potentials reveals us that the Ohm equations for 
the classical Hall effect:
\begin{equation}
j_m\,=\,{{e^2\,n}\over{m}} a_m\,+\,k \epsilon^{mn} \dot{a_m}
\end{equation} 
where $j_m$ is the probability current.
The equation above in the appropriate gauge $a_m\,=\,E_m\tau$ for low magnetic fields, allow us to write the right Ohm equations for the classical Hall effect
\be
j_m =\sigma_L E_m\,+\,\sigma_H \epsilon_{mn}E_n\,\,\,\,\,,
\ee
where k can be immediately understood like the Hall conductivity $\sigma_H$
\section{Symplectic Algorithm and Constraints}
Let us make a short review about some of the main concepts of the Fadeev-Jackiw-Symplectic\cite{FJ,BaWo} algorithm.
Because this one will be fundamental for our results and considerations here.

We consider firstly a Lagrangian that is first order in time derivatives, if the original
Lagrangian has not this form, we can introduce the auxiliary fields and change it to first order.
\be
L^{(0)}\,=\,a^0_{i}\partial_t \phi_i\,-\,U^{(0)}(\phi,\partial \phi)
\ee
where $\phi_k$ are fields belonging the symplectic set.  The symplectic 2 form is defined as:
\be
f_{ij}(x-y)\,=\,{{\delta a_j(y)}\over{\delta \phi_i(x)}}\,-\,{{\delta a_i(x)}\over{\delta \phi_j(y)}}.
\ee
If it is not singular we can calculate the Dirac brackets and define the commutator of quantum theory,
\be
\{A,B\}_D\,=\,(f^0)^{-1}_{ij}\delta^2(x-y)
\ee
However if the 2 form is singular is necessary to calculate the zero modes that satisfy $f^{0}_{ij}v^{\alpha}_i\,=0\,$
and the corresponding constraints:
\be
\label{zeromode}
\Omega^{\alpha}\,=\,\int dx {\tilde v_{\alpha}(x)} {{\delta}\over{\delta\phi(x)}}\int dy U(y)=0 ,
\ee
the constraints are usually introduced in a new kinetic sector of the Lagrangian.
\be
L^{(1)}\,=\,L^{(0)}\,+\,\dot{\lambda}^{\alpha}\Omega^{\alpha}.
\ee

Now our first approach will be to apply the symplectic algorithm to the
model (\ref{Action})rewriting the action regarding the terms with the first derivative on the time. \bea
S^{(0)}&=&\,{1\over{8\pi}}\int dt \quad \Biggl[\int_{\Sigma}
i\hbar \psi_i^\ast
\partial_t \psi^i\,-\,k \, \int_{M}\epsilon^{mn}a_m \partial_t a_n \nonumber \\
&+& U_{Sch}(\psi_i\,,a_m)\,+\,U_{CS}(a_m)\Biggr] \quad \eea where
$U_{Sch}$ and $U_{CS}$ are representing the Schroedinger and Chern-Simons symplectic potential sectors.
The set of symplectic variables under consideration is \be
\Xi^{(0)}\,=\,(\psi_i,\,\psi_i^\ast,\,a_0,\,a_m) \ee
The (degenerated) zero order symplectic matrix is easily found to be
\be f^{(0)}_{ij}(x\,-\,y)\,=\,{{1}\over{8 \pi}}\left(
\begin{array}{cccc}
0            &i \hbar \delta_{ij}  &   0                       & 0 \\
i \hbar\delta_{ij}  &  0           &   0                      & 0  \\
0            &  0           &  0    & 0 \\
0   & 0  & 0 & 2\,k \, \epsilon^{mn}
\end{array} \right)
\delta^2(x\,-\,y). \ee 
The symplectic tensor is singular and the unique zero-mode for this matrix is $\tilde v\,=\,(0,0,v^{A_0},0)$,
where the function $v^{a_0}$ is totally arbitrary.  The zero-mode condiction, Eq.(\ref{zeromode}) select the following constraint
\bea
0&\approx&\int dx {\tilde v^{a_0}(x,t)} {{\delta}\over{\delta a_0(x,t)}}\int dy U(y,t) \nonumber \\
0&\approx&\int dx {\tilde v^{a_0}(x,t)} [k
\epsilon^{mn}\partial_m a_n(x,t)\,-\,e\psi_i^\ast(x,t) \psi^i(x,t)]
\eea
since  $v^{a_0}(x,t)$ is an arbitrary function the constraint is, \be
\label{constraint} \Omega_{a_0}\,=\,k
\epsilon^{mn}\partial_m a_n\,-\,e\psi_i^\ast \psi^i\approx 0 \ee 
is in fact an important relation rising from the model
because it permits, if integrated, the direct access to the well
known relation between the Hall conductivity $\sigma_H$ and the
magnetic field $B$. \be \sigma_H\,=\,{ne\over B}.\ee 
where
\begin{equation}
n\,=\,s^{-1} \int da \psi^{\ast}\psi
\end{equation}
is the global density of charge carriers in a sample area $s$.
Subsequently is necessary now to deform the kinetic sector of action by including
the constraint and reevaluate the symplectic matrix \cite{BaWo}.
\be
L^{(0)}\rightarrow L^{(1)}\,=\,L^{(0)}\,+\,\dot{\lambda}^{a_0}\Omega^{a_0}
\ee
This standard procedure show us a new singular structure where the gauge has not
been fixed yet.  Usually the standard procedure (Weyl gauge) consists to make $a_0\,=\,0$,
and the new set of symplectic variables is now \be
\Xi^{(1)}\,=\,(\psi_i,\,\psi_i^\ast,\,\,a_m,\lambda)\ee
the symplectic tensor will allow us to find easily a matrix picture where
in its elements. Now we are able to construct a non singular first iterated symplectic
matrix \be f^{(1)}_{ij}(x\,-\,y)\,=\,{{1}\over{8 \pi}}\left(
\begin{array}{cccc}
0            &i \hbar \delta_{ij}  &   0                       & e \psi_i^\ast \\
i \hbar\delta_{ij}  &  0           &   0                      & e \psi_i  \\
0            &  0           &2\,\sigma_H\, \epsilon^{mn}            & 0 \\
e \psi_i     & e \psi_i^\ast  & 0 & 0
\end{array} \right)
\delta^2(x\,-\,y), \ee its inverse can be easily evaluated, \be
[f^{(1)}_{ij}]^{-1}(x\,-\,y)\,=\,8 \pi \left(
\begin{array}{cccc}
i \hbar\,\delta_{ij} {{\psi_i}\over{\psi_j^*}}            & i \hbar  \delta_{ij}  &   0  & {{1}\over{e\psi_j^*}} \\
i \hbar  \delta_{ij}  &      i \hbar\,\delta_{ij} {{\psi_i^*}\over{\psi_j}}       &   0  & {{1}\over{e\psi_j}}   \\
0            &  0                           &    {{\epsilon_{mn}}\over{2\sigma_H}}          & 0 \\
{{1}\over{e\psi_i^*}}          &  {{1}\over{e\psi_i}}           &
0 & {{-\,i \hbar \, \delta_{ij}}\over{e^2\,\psi_i^*\,\psi_j}}
\end{array} \right)
\delta^2(x\,-\,y), \ee
we can read the Dirac brackets to gauge field $a_m$. \be \label{brackets}
\{a_m(x,t)\,,\,a_n(y,t)\}_D\,=\,{4 \pi 
\over{\sigma_H}}\epsilon_{mn}\delta^2(x\,-\,y)\hspace{1cm}\footnote{$x,y\in\,\Sigma$}.\ee

\section{Noncommutative Extension}
Recently, there have been a great deal of interest in noncommutative field.  The main was perceived when it was noted that noncommutative spaces naturally arise in perturbative string theory with a constant background magnetic fieldin the presence of D-branes.  For this limit the dynamics of D-branes can be described by a noncommutative gauge theory \cite{Douglas,Witten}. 

It is known that in a quantized system of particles submitted to a strong magnetic field the noncommutative space arises naturally.  It is also know that a system of electrons moving in a strong magnetic field  we can research several interesting properties.  Then the possibility of new and interesting contributions is a good reason to study a noncommutative extension of the model analised before.  Our proposal now is to implement such noncommutative extension of
the model managing the two sectors adequately.   We are interested in understanding how the constraints of the theory 
should be affected by a non commutative contribution to the Schroedinger and
Chern-Simons sectors.  We must inquire about a possible
new and adjusted constraint expression, it probably will connect the Hall
conductivity, in a noncommutative sense, with the magnetic field
and $\theta$.  We can construct a more general action
where the Chern-Simons sector have been rewritten in a
noncommutative sense implementing a well known Moyal-Weyl product \cite{Moyal} of fields.  In this formalism the fields are defined as functions of the phase space variables  with the product of two fields given by \be  \Phi_a(x) \star 
\Phi_b(y)\,=\,exp{\big({i\over2}\theta^{\mu \nu}\partial^x_{\mu}
\partial^y_{\nu}\big)\Phi_a(x) \Phi_b(y)}\vert_{x=y},\ee
where $\theta^{\mu \nu}$ is an antisymmetric and real constant
which characterize the noncommutative nature of fields \be [
\Phi(x)_a\,,\, \Phi(y)_b]\,=\,i\theta^{\mu \nu} \ee so the action for our noncommutative theory can be written by
modified expression \bea\label{ncation}
S^{(0)}_{NC}\,=\,&=&{1\over{8\pi}}\int dt \quad
\Biggl[\int_{\Sigma}  \Psi^{\ast}\star[i\hbar
\partial_t\,-\,{1\over{2m}}(i\hbar
\partial_n\,-\,e \star A_n)^2\,+\nonumber \\
&-&e \star A_0]\star  \Psi\,+\,h.c.\,-\,k \int_M \epsilon^{\alpha
\beta \gamma}A_{\alpha}\star  F_{\beta \gamma}\Biggr]
\quad \eea
\newpage
or in its standard form
\bea\label{ncation}
S^{(0)}_{NC}\,=\,&=&{1\over{8\pi}}\int dt \quad
\Biggl[\int_{\Sigma}  \Psi^{\ast}\star[i\hbar
\partial_t\,-\,{1\over{2m}}(i\hbar
\partial_n\,-\,e \star A_n)^2\,+ \nonumber \\
&-&e \star A_0]\star  \Psi\,\,+\,h.c.\,-\,k \int_M \epsilon^{\alpha
\beta \gamma}(A_{\alpha}\star \partial_{\beta} A_{\gamma}\,-\,{{2i}\over{3}} A_{\alpha}\star A_{\beta}\star A_{\gamma})\Biggr]
\quad \eea{}
The explicit form of the Seiberg-Witten maps connecting the gauge fields was presented in \cite{Witten}.
\bea
\Psi_i&=&\psi_i\,-\,{1\over2}\theta^{mn}a_m\partial_n \psi_i \nonumber \\
A_i&=&a_i\,-\,{1\over2}\theta^{mn}a_m(\partial_n a_i\,+\,F_{ni})
\eea
For our aim we are considering the Moyal expansion to
the first order on $\theta$ \be  \Phi_a (x)\star 
\Phi_b(y)\,=\,\Phi_a(x) \Phi_b(y)\,+\,{i\over2} \theta^{\mu \nu}
\partial_{\mu}\Phi_a(x)
\partial_{\nu}\Phi_b(y). \ee
When expanded to the first order on $\theta$ in the Moyal product
and after some algebra we can easily show that the action becomes

\bea S^{(0)}_{NC}&=&\,{1\over{8\pi}}\int dt \quad
\int_{\Sigma} \Biggl[i\hbar  \Psi^*\partial_0  \Psi\,-\,{{\hbar^2}\over{2m}}\Psi^{\star}\partial^2\Psi\,+\,
{{ie \hbar}\over{2m}}(\Psi^{\star}\partial^n A_n \Psi\,+\,\Psi^{\star}A_n \partial^n \Psi)\,+\,\nonumber \\
&-&{{e\hbar}\over{4m}}\theta^{\mu\nu}(\partial_{\mu}\Psi^{\star}\partial^n A_n\partial_{\nu}\Psi\,+\, 
\Psi^{\star}\partial_{\mu}\partial^n A_n \partial_{\nu}\Psi)\,-\,{{e^2}\over{2m}}\Psi^{\star}A_n A^n \Psi \,+\nonumber \\
&-&{{i e^2}\over{4m}}\theta^{\mu\nu}\Psi^{\star}\partial_{\mu}A_n A^n \partial_{\nu}\Psi\,+\,e\Psi^{\star}A_0\Psi\,-\,{{i e}\over{2}}\theta^{\mu\nu}\partial_{\mu}\Psi^{\star}A_0 \partial_{\nu}\Psi \,\Biggr]+ \Biggl[ h.c. \Biggr]\nonumber \\
&+&{k\over{8\pi}}\int dt \quad
\int_{M} \Biggl[\epsilon^{mn}A_m \partial_0 A_n\,+\,2\epsilon^{mn}A_0\partial_m A_n\,+\,A_0\epsilon^{mn}\theta^{\mu\nu}\partial_{\mu} A_m \partial_{\mu}A_n\Biggr]. \eea
We can easily identify two different sectors in the action they are,
\begin{equation}
S^{(0)}_{NC}\,=\,{1\over{8\pi}}\int dt \quad
\Biggl[\int_{\Sigma}  a_{\Psi}\dot{\Psi}\,+\,U_{\Psi}\,+\,\int_{M}  a_{A_n}\dot{A_n}\,+\,U_{A_n}\Biggr],
\end{equation}
where the action was written again in the first order for the time derivative,
and the two sectors of the model were separated on purpose, and
the variables to be considered on the symplectic approach are
$\Xi^{(0)}\,=\,( \Psi_i, \Psi_i^*,A_0, A_m)$.  We will
have now  to mount the zero order symplectic tensor.  Its form can be easily found,
\be f^{(0)}_{ij}(x\,-\,y)\,=\,{1\over{8
\pi}}\left(\begin{array}{cccc}
0            &-i\hbar\delta_{ij}  &   0       & 0 \\
i\hbar\delta_{ij}  &  0                 &   0             & 0 \\
0            &  0                 &   0             & 0 \\
0            &  0                 &   0       & 2k
\epsilon^{mn}
\end{array} \right)
\delta^2(x\,-\,y).\ee Obviously the matrix is singular, so new constraints shall be obtained.
But now we could inquire about the possible noncommutative nature for the
constraint like \be \Omega_{\theta}\,=\,\Omega(A_m,\Psi,\theta) \ee
where we are considering this new constraint as a combination of
the constraint (\ref{constraint}) written for the original 
model plus a new and interesting noncommutative contribution. The
eigenvector  with zero eigenvalue is $\tilde \nu_{A_0}^{0}$ and
formally we may write the explicity constraint expression 
\bea
0&\approx&\int dx {\tilde v^{a_0}(x,t)} {{\delta}\over{\delta A_0(x,t)}}\Biggl[\int_{\Sigma}
eA_0(y,t)\psi_i^\ast(y,t) \psi^i(y,t)\,+\,\int_{M}(2k\epsilon^{mn}A_0(y,t)\partial_m A_n(y,t)+\nonumber \\
&-&{k\over3}A_0(y,t)\epsilon^{mn}\theta^{\mu\nu}\partial_{\mu}A_m(y,t) \partial_{\nu}A_n(y,t))\Biggr]
\eea
the eigenvector $\tilde \nu_{A_0}^0$ is an arbitrary function so we obtain the
exact form for the constraint, which is, \be
\Omega_{\theta}\,=\,2e  \Psi_i^*  \Psi^i\,-\,2k
\epsilon^{mn}\partial_m A_n\,-\,k
\epsilon^{mn}\theta^{ij}\partial_i A_m  \partial_j 
A_n\,\approx \,0 \ee 
This is the main and more important result that we could obtain using the only the Fadeev-Jackiw
aproach to constrained system.  The noncommutative theory considered above is
presenting a new and interesting form of correcting the old
constraint, showing us a new physical sector. By means of this new
constraint relation we can express the  conductivity
$\sigma^{\theta}_H$ in a straightforward and natural way with noncommutative
correction. If now we integrate the constraint relation and
consider \footnote{$B=\epsilon^{mn}\partial_m A_n$, $e  \Psi^*
 \Psi=j_0$ and $(\nabla \times
A)^2=\theta\,\epsilon^{ij}\epsilon^{mn}\partial_iA_m
\partial_j A_n$} some basic relations, the expression for the
conductivity can be written as, \be \label{cond}
\sigma^{\theta}_H\,=\,{{n\,e}\over{ B\,+\,{{1}\over2}} \theta
 B^2}\,=\,\sigma_H f(\theta,B)\ee Subsequently we need to include the constraint in the
model, enlarging the kinetic Lagrangian sector.  This will be done
by including a new variable to the system, using the consistency
condition like a Lagrange multiplier ($\lambda$). The twice
iterated Lagrangian is obtained by bringing the primary constraint
into canonical part of $L^{(0)}$  \be L^{(1)}\,=\,L^{(0)}\,+\,\dot
\lambda \Omega_{\theta},\ee through this new Lagrangian it is
possible to construct the following new symplectic tensor with the
set of variables \be \label{xi}
\Xi^{(1)}\,=\,( \Psi_i, \Psi_i^*,A_0, A_m,\lambda). \ee
But once more we are dealing with a singular matrix because the
model still has a gauge symmetry.  Let us choose again the same
gauge fixed used before (Our choice is the condition $A_0\,=\,0$,)
where we are retaining the real degrees of freedom of
electromagnetic fields, we can eliminate this additional
obstruction and write the correct two form symplectic tensor, 
belonging to the symplectic set (\ref{xi}). So its basic form can
be written as,
 \be f^{(1)}_{ij}(x\,-\,y)\,=\,{1\over{8
\pi}}\left(\begin{array}{ccccc}
0                  &-i\hbar\delta_{ij}  &   0       & 0   & e  \Psi_i^*  \\
i\hbar\delta_{ij}  &  0                 &   0       & 0   & e  \Psi_i \\
0                  &  0                 &   0       &
-2\sigma_H \epsilon^{mn} & -2\sigma_H \epsilon^{mn} \partial_m \\
-e  \Psi_i^* & -e  \Psi_i & 0 & 2\sigma_H \epsilon^{mn} \partial_m &  0
\end{array} \right)\delta^2(x\,-\,y) \ee
This is a nonsigular matrix and the corresponding inverse is easily obtained by a direct and simple calculation, and we can identify for this tensor the 
same structure for the Dirac brackets found before. \be \label{brackets}
\{A_m(x,t)\,,\,A_n(y,t)\}_D\,=\,{4 \pi 
\over{\sigma_H}}\epsilon_{mn}\delta^2(x\,-\,y) \ee
for the gauge fields.
\newpage
\section{Discussion}

We have considered the classical and integer Quantum Hall effect based on a semi classical action with two sectors (Schroedinger plus Chern-Simons) and we have applied the Fadeev-Jackiw formulation to research the constraints of the model.
Firstly we have analyzed the ordinary model and reproduced the standard results for the Hall conductivity.  Subsequently we considered a noncommutative extension and repeated the treatment. In this way we showed that it is possible to obtain a new  and corrected Hall conductivity expression only by evaluating the constraint expression, we must observe that the new result is a natural consequence implicit in the model and arose like a strong relation connecting the main physical parameters of the theory showing a new and unexpected meaning of the constraints, in fact an interesting estimative to $\sigma$ could be found when we considered this approach. The relation (\ref{cond}) expresses the interesting new dependence of $\sigma$ with the noncommutative parameter $\theta$ it is evident that the first piece dominates the behavior showing us the weak influence of the second piece.  An important observation must be done about the agreement of our relation with that one presented by Susskind when he has proposed that noncommutative Chern-Simons theory is a better description of fractional quantum Hall states, reproducing the detailed properties of the quasiparticles \cite{Sus}.  The both relations have showed the right dependence
\be
\theta\,\sim \,{1\over B}.
\ee
The noncommutative potential sector had an important participation to generate the
constraint relation, in fact this new noncommutative dynamics
deformed the usual model and adds some nontrivial terms leading to
$O(\theta)$ corrections. In this sense, we can suggest a new Hall
conductivity where an effective magnetic field arises naturally
\bea
B_{eff}\,=\, B\,+\,{1\over2}\theta  B^2 \nonumber, \eea
obviously in the limit $\theta \rightarrow 0$ we
re obtain the usual results of original theory.

In addition there is some speculations about the connection between the noncommutative sector with the impurities inherent to this experiment this would reinforce our result as a new and corrected expression for the Hall conductivity obtained only by a treatment of a constrained system is a different way to approach old result and to obtain new ones. 

We can still enquire that more contributions in higher order of
$\theta$ could be obtained from the constraint relation if we had
considered higher $\theta$ terms in the Moyal expansion, these new
constraints could have as many terms as necessary. Another
important aspect is that the usual Dirac brackets retained the
same form showing us that symmetry breaks do not occur when the
noncommutative transition is done.
\section*{Acknowledgments}
I am extremely grateful to Prof. J. Abdalla Hela\"yel-Neto for
discussions and the final revision of the paper.  Some short discussions
changed deeply the main idea of this text.
\\
This work is supported by FAPERJ-RJ, Brasilian Agency of Research

\end{document}